\documentclass[11pt,conference,letterpaper]{IEEEtran}

\usepackage{etex} 
\usepackage[utf8]{inputenc}
\usepackage{cmap}  
\usepackage[T1]{fontenc}
\usepackage[english]{babel}
\usepackage{cite}
\usepackage{hyperref}
\hypersetup{final,
	        plainpages=false,
	        pdfpagelabels,
	        pagebackref,
	        colorlinks=true,
	        linkcolor=blue,
	        urlcolor=red,
	        citecolor=blue,
	        pdfpagemode=UseOutlines,
	        pdfstartview=FitH,
	        pdfborder={0 0 0}}
\usepackage{url}
\usepackage[per-mode=symbol,bracket-unit-denominator=false,detect-all,load-configurations=binary,binary-units]{siunitx}
\DeclareSIUnit\loc{LOC}
\DeclareSIUnit\permil{\textperthousand}

\usepackage{booktabs}
\usepackage{graphicx}
\usepackage{multirow}
\usepackage{tikz}
\usetikzlibrary{arrows,positioning,shapes,topaths,calc,fit,backgrounds,matrix,shadows,automata,patterns,decorations.pathmorphing,decorations.pathreplacing,decorations.text,circuits.logic.US,trees,mindmap}
\usepackage{microtype}

\begin{document}
\title{The Need for a Versioned\\ Data Analysis Software Environment}

\author{\IEEEauthorblockN{Jakob Blomer, Dario Berzano, Predrag Buncic, Ioannis Charalampidis,\\ Gerardo Ganis, George Lestaris and Ren\'e Meusel}\IEEEauthorblockA{European Organization for Nuclear Research\\CERN PH-SFT\\1211 Gen\`eve 23, Switzerland\\Email: jblomer@cern.ch}}

\maketitle

\begin{abstract}
	Scientific results in high-energy physics and in many other fields often rely on complex software stacks. 
	In order to support reproducibility and scrutiny of the results, it is good practice to use open source software and to cite software packages and versions. 
	With ever-growing complexity of scientific software on one side and with IT life-cycles of only a few years on the other side, however, it turns out that despite source code availability the setup and the validation of a minimal usable analysis environment can easily become prohibitively expensive.
	We argue that there is a substantial gap between merely having access to versioned source code and the ability to create a data analysis runtime environment.
	In order to preserve all the different variants of the data analysis runtime environment, we developed a snapshotting file system optimized for software distribution.
	We report on our experience in preserving the analysis environment for high-energy physics such as the software landscape used to discover the Higgs boson at the Large Hadron Collider.
\end{abstract}

\section{Introduction}
\label{sec:intro}
Complex scientific instruments, such as the particle detectors at the Large Hadron Collider (LHC), require an extensive scientific software landscape~\cite{lhc09}.
The raw data volume from the LHC particle detectors exceeds \SI{100}{\peta\byte} and it is expected to reach the exabyte scale in the next few years.
At the same time, the data are not self-explanatory and the true physics information is extracted in a computationally intense process.
For instance, as a result of a particle collision, up to 100 million detector channels are evaluated to reconstruct the location of a collision, the trajectory of particles that traverse the detector, and their charge and momentum; clusters of particles are identified and separated, particle-types are deducted, and error estimations due to detector inefficiencies are calculated.
In the course of data analysis, statistical methods and machine learning algorithms separate a background of physics processes from a signal that allows rare or new physics processes to be studied.
Data analysis and detector development also require simulation studies that produce an amount of data in the same order of magnitude as the data taken from a detector.
The physics processes affecting particles traversing the detector material are described by a plethora of different simulation algorithms.
Overall, the scientific software suite that was used in the discovery of the Higgs boson~\cite{higgsatlas12,higgscms12} comprises several millions of lines of code, which is comparable to the complexity of an office suite or an operating system~\cite{loc13}.

\section{Common Problems}
\begin{figure}
	\begin{center}
		\resizebox{0.35\textwidth}{!}{\begin{tikzpicture}
	\tikzset{
		block/.style={thick,rounded corners,draw,align=center, fill=white, drop shadow, minimum width=4cm, minimum height=1.5cm, top color=white},
	}
	\colorlet{colexternal}{red!50!black!20}
	\colorlet{colhep}{green!50!black!20}
	\colorlet{coluser}{blue!50!black!20}
	
	\node[label={[rotate=20,font=\bf,red!75, label distance=2mm]right:20 MLOC}, block, bottom color=colexternal] (os) at (0,0) {Grid Libraries\\System Libraries\\OS Kernel};
	\node[label={[rotate=20,font=\bf,green!75!black!75, label distance=2mm]right:5 MLOC}, above=4mm of os.north,block,bottom color=colhep,double copy shadow] (root)  {High Energy Physics\\Libraries};
	\node[label={[rotate=20,font=\bf,green!75!black!75, label distance=2mm]right:4 MLOC}, above=4mm of root.north,block,bottom color=colhep,double copy shadow] (framework) {Experiment\\Software Framework};
	\node[label={[rotate=20,font=\bf,blue!75!black!75, label distance=2mm]right:0.1 MLOC}, above=4mm of framework.north,block,bottom color=coluser,double copy shadow] (analyse) {Individual \\Analysis Code};
	
	\draw[very thick,->] (5.5,-0.5) -- node[very near start,rotate=90,red,yshift=-3mm] {stable} node[very near end,rotate=90,blue,yshift=-3mm] {changing} (5.5,6);
	
\end{tikzpicture}}
	\end{center}
	\caption{The high-energy physics software stack of an LHC experiment.}
	\label{fig:swstack}
\end{figure}
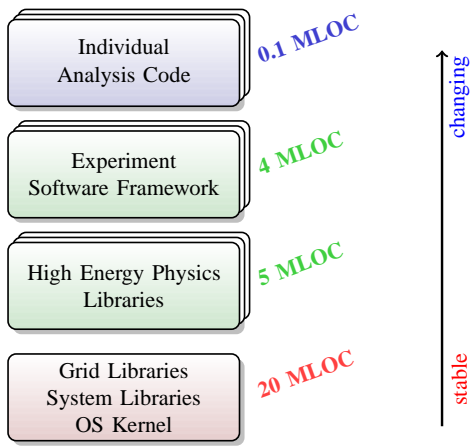

The typical high-energy physics software stack is shown in Figure~\ref{fig:swstack}.
At the infrastructure level, general-purpose libraries and tools are used, while at higher levels the software is highly specialized to address particular physics questions.
In comparison to commodity software of similar scale, we see the following issues that make the deployment of large-scale scientific software more difficult:
\begin{enumerate}
	\item In order to ensure correctness, the software stack does not only undergo unit tests but larger changes require a lengthy validation of the physics output (so-called \emph{physics validation}).
		In the process of validation, a particular software version is used to process a large data sample and the results are compared to simulation studies and previously validated versions.
		Validation requires both deep knowledge of the physics as well as deep knowledge of the software.
		Thus validation involves tens of people and it can take several weeks or even a few months. 
		Therefore, the specialized scientific software is only available for a small number of operating system versions and compilers.
	\item The more specialized the software, the closer becomes the number of users as compared to the number of developers. 
		\SI{100}{\kilo\loc} of ``individual analysis code'' (Figure~\ref{fig:swstack}) has a ratio of users per developer of 1.
		Subtle deployment problems, such as build system errors, hidden library dependencies, or missing ``includes'', however, require a large user base to be discovered.
		Moreover, only a large user base provides the incentive for the mundane tasks involved in fixing deployment problems.
	\item Scientific software changes quickly.
		Table~\ref{tab:releases} shows that major releases of the software frameworks of large LHC experiments are created weekly or bi-weekly.
		Smaller patches and tags of individual analysis code are published and deployed on a daily basis.
		For the daily work of researchers, scientific software is a moving target and the result of a lengthy installation effort can easily be an outdated system.
\end{enumerate}

The typical first step in validating and refining a data analysis is to replicate the data analysis environment of the original publication.
The effort to create a usable data analysis environment is substantial.
The researcher needs to setup the same operating system, compilers, tools, and libraries that were used in the original analysis which might have become outdated and unavailable in the meantime.
Not all dependencies are clearly specified but they are often untracked or implicitly specified as part of possibly unversioned build system scripts.
This is not a particular problem of scientific software but rather a problem of scale; fighting such problems is well-known in the communities that maintain Linux software repositories.

\begin{table}
	\begin{center}
		\begin{tabular}{lrr}\toprule
				& {\bf\centering Releases}	& {\parbox{3cm}{\bf\centering Shared libraries\\ and plug-ins\\ per release}}\\\midrule
		ATLAS	& 19					& 3900\\
		CMS		& 40					& 2200\\
		ALICE	& 49					& 210\\\bottomrule
		\end{tabular}
	\end{center}
	\caption{LHC experiment release statistics for 2013.
		The LHCb experiment is missing because its framework is not released as a whole but as individual sub-projects.}
	\label{tab:releases}
\end{table}

Unfortunately, modern software engineering practices amplify the problem of deployment.
Rather than building monolithic systems, modern software engineering fosters small, pluggable buildings blocks that can be connected and configured to act as a coherent system.
The software for the LHCb experiment at the LHC, for instance, comprises some 40 sub-projects with their own development teams and release plans.
The knowledge about the inter-dependencies of these projects is not entirely automatically tracked.
Some of the knowledge about which project versions need to be used together in which configuration is only captured in the role of software librarians (human beings) or spread over numerous wikis and web pages.

\section{A Time Machine for the Data Analysis Environment}
In an attempt to capture an entire, usable data analysis environment, we started to look into hardware virtualization.
Unfortunately, virtualization introduces a new problem, that of keeping track and distributing hard disk images.
With many gigabytes of software that changes daily and that needs to be deployed across a world-wide distributed computing infrastructure, the traditional image building and deployment process becomes prohibitively expensive.

To solve the problem, we developed CernVM-FS, a distributed file system that is optimized for software delivery~\cite{cvmfs11}.
In CernVM-FS, software is installed only once on a central server.
Data are distributed to clients through HTTP using a global network of caching web proxies.
The file system is designed for the characteristics of software: small files, high meta-data request rate, a high level of redundancy, and a small working set of files for every particular client.
CernVM-FS is based on content-addressable storage and hash trees which facilitates caching, provides data de-duplication, and allows for file system snapshots and versioning.

Due to its build-in versioning, CernVM-FS can also be seen as a ``time machine'' for the data analysis environment.
Any file system snapshot can be mounted on the client side, so all previous states of the data analysis environment are automatically preserved and easily accessible.
By placing the scientific software, compilers, tools, and operating system binaries on CernVM-FS, all that remains in a virtual machine image is the operating system kernel and the file system client~\cite{ucernvm14}.
Image building and deployment become trivial.

While originally developed for cloud computing environments, we seemed to strike a chord in the grid community.
For the LHC experiments with their world-wide distributed grid computing infrastructure, we managed to reduce the time between building a new software release and the rollout from several days to less than one hour.
As a result there is now a big increase in the number of scientific users in high-energy physics and in various other fields~\cite{cvmfsother14,cvmfsother214}.

\section{Open Issues and Future Work}
In the same way version control systems helped to keep large code bases manageable, we believe that a versioning system tailored to the needs of software binaries can help to keep ever-changing data analysis environments manageable.
Our efforts allow for snapshotting and deployment of the data analysis environment.
Several aspects of version control systems~\cite{vcs75}, however, remain unsolved for software binaries.
In particular, we are currently unable to answer the question what has changed and why between any two file system snapshots.
This is an essential question when it comes to connecting preserved software components with new software components.
For example, researchers might want to compare data from contemporary simulation algorithms with those of a preserved analysis of historical data.

In addition to the possibilities a snapshotting file system can offer, we think it would be beneficial if future generations of automatic software build systems not only executed build scripts but also created a ``build recipe'' that concisely describes the conditions and actions necessary to create a data analysis environment.
Such mechanisms already exist in a simplistic manner, for instance the automatic dependency detection of package managers or build systems that burn the build environment and compile options into the binary.

Finally, more efforts are necessary to not only capture the environment of a single computer but to capture a data analysis environment that depends on a complex distributed infrastructure.
Critical external services may for instance include databases, distributed storage, directory and authentication servers.
When it comes to the long-term preservation of the ability to analyze historical data, critical external services must be reduced to a minimum and they should connect through long-lived and lean interfaces to the scientific software.

\section*{Acknowledgments}
We would like to thank Benedikt Hegner, Andreas Pfeiffer, and Alessandro De Salvo for their input on the LHC experiment software release statistics.

\bibliographystyle{IEEEtran}
\bibliography{bibliography}

\end{document}